\RequirePackage{fix-cm}
\documentclass[twocolumn]{svjour3}          

\smartqed  

\usepackage{graphicx}
\usepackage{url}

\usepackage{hyperref}
\hypersetup{breaklinks}
\usepackage{breakurl}
\usepackage{latexsym}
\usepackage{newtxmath}
\usepackage{amsmath}
\usepackage{tabulary, booktabs}
\usepackage{colortbl}
\usepackage{multirow}
\usepackage[numbers]{natbib}
\usepackage{natbib,stfloats}
\usepackage{mathrsfs}
\usepackage{graphicx}
\usepackage{dirtytalk}
\usepackage{url}

\journalname{ArXiv}
\begin{document}
\authorrunning{Sarah Radway et al.}
\title{Beyond LunAR: An augmented reality UI for deep-space exploration missions 
}

\titlerunning{AR UI for Deep Space Missions}        
\author{Sarah Radway, Anthony Luo, Carmine Elvezio, Jenny Cha, Sophia Kolak, Elijah Zulu, Sad Adib }

\institute{Sarah Radway, Anthony Luo, Carmine Elvezio, Jenny Cha, Sophia Kolak, Elijah Zulu, Sad Adib \at
              Department of Computer Science \at
              Columbia University \at
              New York, NY 10027, USA \at
              \email{sar2236@columbia.edu}           
}

\date{}

\maketitle

\begin{abstract}
As space exploration efforts shift to deep space missions, new challenges emerge regarding astronaut communication and task completion. While the round trip propagation delay for lunar communications is 2.6 seconds, the time delay increases to nearly 22 minutes for Mars missions \cite{LOVE2013}. This creates a need for astronaut independence from earth-based assistance, and places greater significance upon the limited communications that are able to be delivered. To address this issue, we prototyped an augmented reality user interface for the new xEMU spacesuit, intended for use on planetary surface missions. This user interface assists with functions that would usually be completed by flight controllers in Mission Control, or are currently completed in manners that are unnecessarily difficult. We accomplish this through features such as AR model-based task instruction, sampling task assistance, note taking, and telemetry monitoring and display. 

\end{abstract}
\section{Introduction}
Human space exploration has always been heavily supported by the people in Mission Control, who communicate information to astronauts through voice communication channels, known as voice loops \cite{PATTERSON1999}.

Astronauts rely on these communications for mission critical information, such as task instructions, telemetry status, or commands in unexpected circumstances. As space exploration moves into deep-space, however, communication delays from Earth can be up to 22 minutes on Mars \cite{LOVE2013}, making the current means of information transfer unsuitable.

As NASA prepares to send astronauts into more remote parts of space for longer periods of time \cite{DUNBAR2020}, there has been significant research into the effects these delays will have on communication between astronauts and Mission Control. Love and Reagan \cite{LOVE2013} find that when delays are introduced, audio communication is hindered by disordered sequences, interrupted calls, impaired ability to provide relevant information, slow response to events, and reduced situational awareness. 

In response to the issues caused by delayed audio transmissions, NASA and commercial partners are developing spacesuits with augmented reality (AR) enabled glass, that allow graphics to be displayed in the astronaut's visual field \cite{mitra2018human}. By utilizing this ability to display graphics in the field of view of the astronauts, these spacesuits can respond to many of the aforementioned problems introduced by deep-space communication. NASA's Exploration Extravehicular Activity Mobility Unit (xEMU) spacesuits will likely include AR enabled glass for the Artemis 2024 Mission and future missions to Mars, aiding astronauts with difficult tasks such as navigation, science sampling, and critical repair tasks in ways previously unimaginable \cite{mitra2018human}. For instance, instead of being \textit{told} how to repair a rover, an astronaut could be \textit{shown} how to preform the repair through 3D graphics overlaid on the vehicle itself similar to Henderson and Feiner \cite{feinerpaper2}.

User interfaces (UIs) must mediate the interaction between the astronaut and this information, making decisions about when and how to show text and/or graphics. In the context of general AR UIs, Olsson et al. \cite{OLSSON2012} demonstrated that certain approaches do not enhance the user's experience. In safety critical situations such as Extra-Vehicular Activities (EVAs), it is of the utmost importance that astronauts have a thorough understanding of instructions and their surroundings. Thus, while AR has the capacity to enrich communication in space, its effectiveness is largely determined by the quality of the UI, which integrates this sensory information into the work environment. 

Because the xEMU spacesuit and its new graphic capabilities have not yet been utilized in a real mission, it is still unclear how graphical elements should be added into the communication loop between Mission Control and astronauts during EVAs. Furthermore, UIs have not yet been developed to cope with the delays between Earth and Mars, which is an important consideration as we enter the era of crewed deep-space exploration. In this paper, we present an AR system designed to address the specific challenges faced by astronauts on deep-space exploration missions. In particular, we aim to alleviate the burden of communication delays, provide a helpful on-demand source of assistance, and facilitate task completion with greater efficacy through hybrid graphics and text instructions. 

Compared to prior work, such as that of Feiner et al. \cite{stevepaper}, our system focuses less on deep-space task management design and implementation, but instead introduces an AR system for EVA assistance for astronauts. Furthermore, our system incorporates elements of AR such as 3D overlaid instructions, like those of D'Souza \cite{dsouza}, which brings about significant user benefits. In contrast to these works, we introduce a voice-based control system for ease of use (given the restrictive nature of spacesuits) and automatic telemetry monitoring and alerting. Our system is specifically designed to increase astronaut independence from MCC, which is especially important given the time-delay of deep-space missions. 

In this paper, we focus on how multimodal systems can be utilized to mitigate the effects of delayed communication in deep space. Our design focuses on minimal physical interaction, intuitiveness, and conciseness, so as not to unnecessarily distract the user. Therefore, we chose a voice-navigation technique to avoid adding to the intense physical strain astronauts face, and implemented an instruction method involving they overlay of AR models on top of the physical environment. Our instruction method aims to eliminate the cognitive strain and confusion of text-only instructions, and instead offer a more intuitive and user friendly combination of text instructions with overlaid AR visualizations. Mindful of the limitations in the way of deep-space communications bandwidth, we established a data flow that reduces the need for task instruction data transmission by at least two orders of magnitude in comparison with a standard approach. Through this project, we hope to contribute to the consideration of the unique needs of astronauts completing lunar and deep-space missions, and propose a solution to a number of the challenges they face.

\section{Background and Related Work}
We begin by describing the necessary background for understanding and evaluating our approach. 
\subsection{Augmented Reality (AR)}
\textit{Augmented Reality (AR)} is an emerging technology which allows computer systems to add virtual content to the visual field of the user \cite{arpaper}. Contrary to \textit{virtual reality (VR)}, which immerses the user in a different environment, AR systems create a layer on top of one's pre-existing visual frame. 
Popular games such as Pok\'emon Go \cite{pokemongo} and Ingress \cite{ingress} are examples of this technology.
\subsection{Deep-Space Exploration}
\textit{Deep-space exploration}, as the name suggests, refers to the exploration of remote regions of outer space. Specifically, these regions must lie outside of lower-earth orbit, which includes missions orbiting around the earth such as the International Space Station, and outside of cis-lunar space, which includes lunar orbiting missions such as the planned Lunar Gateway project \cite{MARS2018}. Examples of deep-space exploration include missions to Mars and Venus. As of now, all deep-space exploration has consisted of crewed missions; however, NASA has shifted its future mission objectives towards manned deep-space exploration, specifically of Mars. 

Significant work has been completed investigating the applications of AR to astronaut task assistance. 
From astronaut training \cite{BRALY2019} to astronaut work support \cite{HELIN2018}, it has been demonstrated that the use of AR improves user experience and productivity. Example implementations include providing system/environment warnings, procedural task assistance incorporating 2D and 3D elements \cite{HELIN2018}, and informational resources \cite{MARKOV2013}. AR is being used on board space missions to assist astronauts, and solve issues that prevent task completion \cite{ramsey_2015}. 

Additionally, the issue of delayed communications accompanying deep-space exploration has been shown to have a significant impact on astronaut performance; NASA astronaut study participants have directly stated that communication delays will represent a massive challenge for long duration missions \cite{KINTZCHOU2016}. Researchers have attempted to mitigate the negative effects of this delay, utilizing techniques ranging from analog testing to behavioral implementations.
Further, Love and Reagan's investigation showed that astronauts preferred to receive non-critical information from Mission Control via text and data communication methods, rather than voice communication methods, when in situations involving significant communication latency \cite{LOVE2013}.
It is important to consider how this general preference can be applied to astronauts' daily tasks in deep-space missions. We were unable to find any academic research examining the application of this preference to astronaut on-board assistance through an AR UI.

\section{Approach}
\subsection{Voice Recognition for User Interface Navigation}
The average astronaut suit has a mass of 127 kg and significantly restricts movement, which makes gestures as simple as scrolling with one's finger highly strenuous \cite{DUNBAR2020}. 
Thus, when it came time to determine how to navigate through our UI, this special condition ruled out many of the more typical methods: namely navigation techniques requiring physical movement.
Astronauts typically communicate with the Mission Control Center (MCC) over voice loops. By building upon this preexisting voice loop framework in the MCC, astronauts would no longer need to halt task completion to provide a status update---flight controllers in Mission Control would be able to access voice data as astronauts work their way through assignments.  

As crewed missions go further into deep space, the communication delays between MCC and on-board crew will become more and more significant, and inevitably increase the level of distraction during the task caused by flight controller interruption \cite{TAYLOR2014}; this ability for a flight controller to passively observe astronaut status could greatly benefit astronaut productivity. 
The use of voice control as a mode of UI navigation in aerospace-related tasks has been thoroughly established and supported in the work of Helin et al. \cite{HELIN2018} and Markov-Vetter and Staadt \cite{MARKOV2013}. Voice control, built upon the preexisting voice loop infrastructure, seems to be the least demanding and most comfortable UI navigation technique that possesses an acceptable accuracy level.

\begin{figure*}[!htb]
\centering
\includegraphics[width=0.75\textwidth]{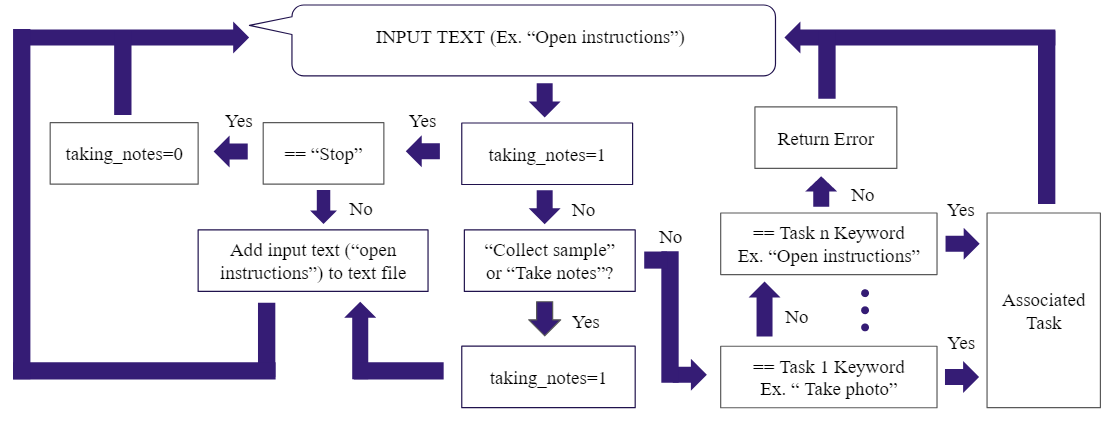}
\caption{Voice Control Basic Outline}   
\label{fig:voiceControl}

\end{figure*}
We use Microsoft Azure Cognitive Speech Services speech-to-text functionality \cite{AzureCogServices}, in order to implement the process control diagram displayed in Figure \ref{fig:voiceControl}. 
We establish a key phrase for each of the UI's basic tasks, which the system identifies when said by the user. 
The UI also displays the most likely relevant key phrase(s) during tasks, to assist the user. 
For example, the UI instructs the user to say the key word \say{next} to scroll to the next instruction. 
The system then recognizes that tag word, and advances the UI to the next step's description. 
This system allows a user to effectively open, close, scroll forwards and scroll backwards through UI elements. 

\indent Additionally, we used voice recognition to assist in tasks involving astronaut note-taking. 
This feature, when activated, records the user's speech to text content on the onscreen notepad display, and stores a file version to send back to the MCC. 
This proves especially helpful for sampling tasks: while harvesting moon rocks on the lunar surface, typically an astronaut would need to record information about the rock samples with a paper and pencil \cite{WONG2017} or the modern equivalent of a tablet and stylus.
The traditional writing approach, aside from being inefficient, would prove uncomfortable for astronauts in a full spacesuit, with limited hand dexterity. 
This speech functionality allows the user to carry out the sampling process without this same need for physical movement; in our implementation, when a rock sample is collected, a virtual notepad was displayed, sequentially asking the user a series of questions, and recording their answers on screen. 
The UI displays the question \say{What color is the sample?}, and the user simply needs to speak to record the color, saying, for instance, \say{gray}. 
When the user has answered the question, and can clearly see that their answer is recorded correctly, they say \say{next}, and can continue working their way through the sample collection, entirely eliminating the need for additional physical burdens. 

While we understand that using a cloud computing service would be infeasible for deep space (for the same reasons as MCC communication), we believe that a majority of our implementation can be replicated without the need for a service like Microsoft Azure.
The vocabulary integral to this UI is very sparse, with fewer than 25 necessary key terms for navigation purposes. 
Beyond that, for sampling tasks, the required vocabulary is also relatively limited. 
Given the relatively bounded expectations for the voice control system, it should be possible to create a system capable of functioning without an Earth-bound computing service, for example by creating a voice control system based on HMMs (Hidden Markov Models), specifically limited in its application and scope, as implemented in Markov-Vetter and Staadt\cite{MARKOV2013}. 

\subsection{Sampling}
The sampling task consists of two main goals: to display and store sampling information and to obtain visual information of the environment.
Given the fact that sampling tasks can be achieved with limited vocabulary, our main focus is to simplify the procedure for the user, as well as to store and display this data in a way that minimized cognitive load and space usage.
To trigger the sampling voice commands, the user says \say{begin sampling}. The procedure is broken down into a sequence of atomic expressions  (true/false) that are triggered by key phrases and words (\say{collect sample}, \say{stop}, \say{exit}). Once the \say{begin sampling} key phrase has been recognized, the system creates an on-device folder for the environment and each sample in order to organize information for both the user and MCC. The system then listens for voice input; the astronaut's observations about the sample along with the date and time are recorded into a text file, which can be displayed in the UI immediately and easily sent to MCC. To obtain visual information, we utilize Unity's \cite{technologies} built in PhotoCapture API, which is integrated with 3D environments, and captures images every 0.5 seconds. We consider both photo and video capture; however our implementation uses just photo capture due to the smaller file size and the fact that the environment does not change enough to require constant surveillance.

\subsection{Telemetry}
When it comes to working in outer space, the immediacy and clarity of alerts is a matter of life or death.
For this reason, we created a telemetry monitoring and reporting system, intended to deal with emergency situations during an EVA. Our system is constantly monitoring and updating telemetry as it flows in, as in the on-board astronaut health analytic system demonstrated in McGregor  \cite{MCGREGOR2013}. 
Upon receiving telemetry data, it compares each value against a standard value range. 
If the value is in a normal range, this information will never be communicated to the user, with the exception of the suit's estimated oxygen, battery, and water time remaining values, which are constantly displayed in the top left corner of the UI. If the value is not in a normal range, however, the system has a series of responses.  

For abnormalities in critical telemetry, such as water or oxygen level, the entire UI will flash a mostly-transparent shade of red so as to not obstruct the user's vision, while ensuring the astronaut understands the urgent need to return immediately to base.
The display warning will continue to appear in red at the top left corner of the UI, making it almost impossible to miss, but once again, not  obstructing the astronauts field of view.
For abnormalities in telemetry considered important, but not immediately life threatening, such as a slight increase in environmental pressure, the system will still display the warning in red at the left corner, however the entire UI will not flash a red hue as the emphasis may serve to distract more than alert. 
In this way, our system attempts to balance emphasis with relevance, ensuring astronauts are always immediately aware when a dangerous situation arises and remain informed of the functionality of their equipment. 

\subsection{Object Recognition and Task Instructions}
Because astronauts have a high cognitive load during an EVA, we aimed to make task instructions as simple and intuitive as possible. 
Therefore we choose to display text instructions in a helmet-fixed manner, meaning that it is in the field of view of the user at all times, and fixed to the coordinate system of the helmet. We accompanied text instructions with an AR overlay on relevant machinery, in order to demonstrate features of the task, similar to that in Markov-Vetter and Staadt \cite{MARKOV2013} and Feiner et al. \cite{feiner1993knowledge}.

To demonstrate the feasibility and accuracy of this system, we implemented instructions for changing a tire on the 
Multi-Mission Space Exploration Vehicle (MMSEV) rover.
An example instruction from this task set may say, \say{Turn the wrench counterclockwise to loosen a bolt that holds the tire in place}; and accordingly, a large red box would be overlaid around the specific bolt, to clarify which bolt the user should loosen, and an arrow would be projected circling around the bolt, indicating to the user which direction they should turn the wrench.

Only two pieces of information are required in order to achieve this overlay-based instruction schema: the 6DOF pose of the target object (in our case, the rover) onto which the instructional models will be projected, and a set of predetermined coordinates of the instructional models in reference to the target object. 

\begin{figure}[!htb]
\centering 
\includegraphics[width=0.45\textwidth]{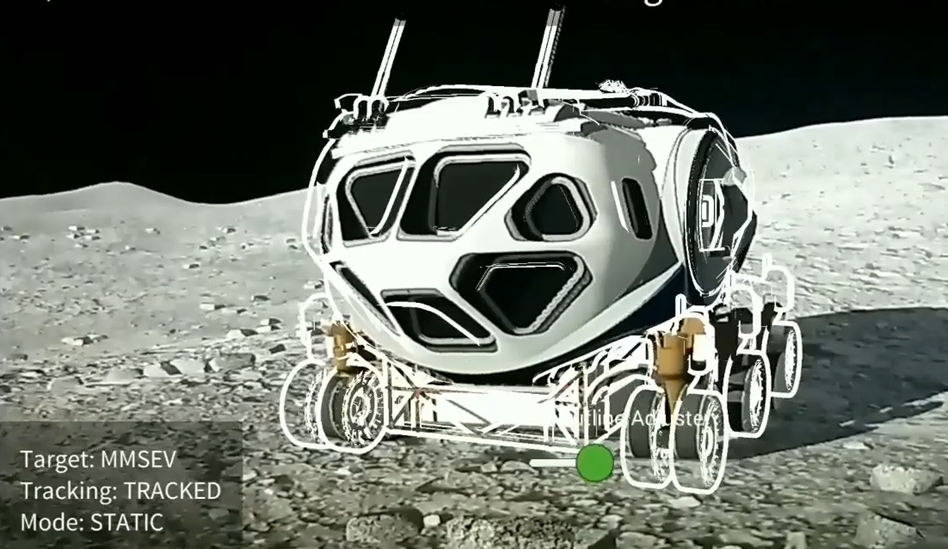}
\caption{Guide View for Tracking of MMSEV Rover}   
\label{fig:rover}
\end{figure}

To meet the first requirement, our system uses the Vuforia Engine \cite{vuforia} in Unity to identify an object and provide its position and orientation. 
We began by establishing a Model Target, a concept in Vuforia representing the base object for use in object and pose recognition, relative to which the augmentations will be placed. In the tire-changing example, the model target would be the rover, as the system is intended to project instructions in reference to the rover as the user goes through the task. 
With Vuforia, a user first aligns the model target with a guide view to begin tracking; to begin the tire-changing task, a user would stand away from the rover, allowing the outline of the guide view to align with the rover as shown in Figure \ref{fig:rover}. 

\begin{figure}[!htb]
\centering
\includegraphics[width=0.45\textwidth]{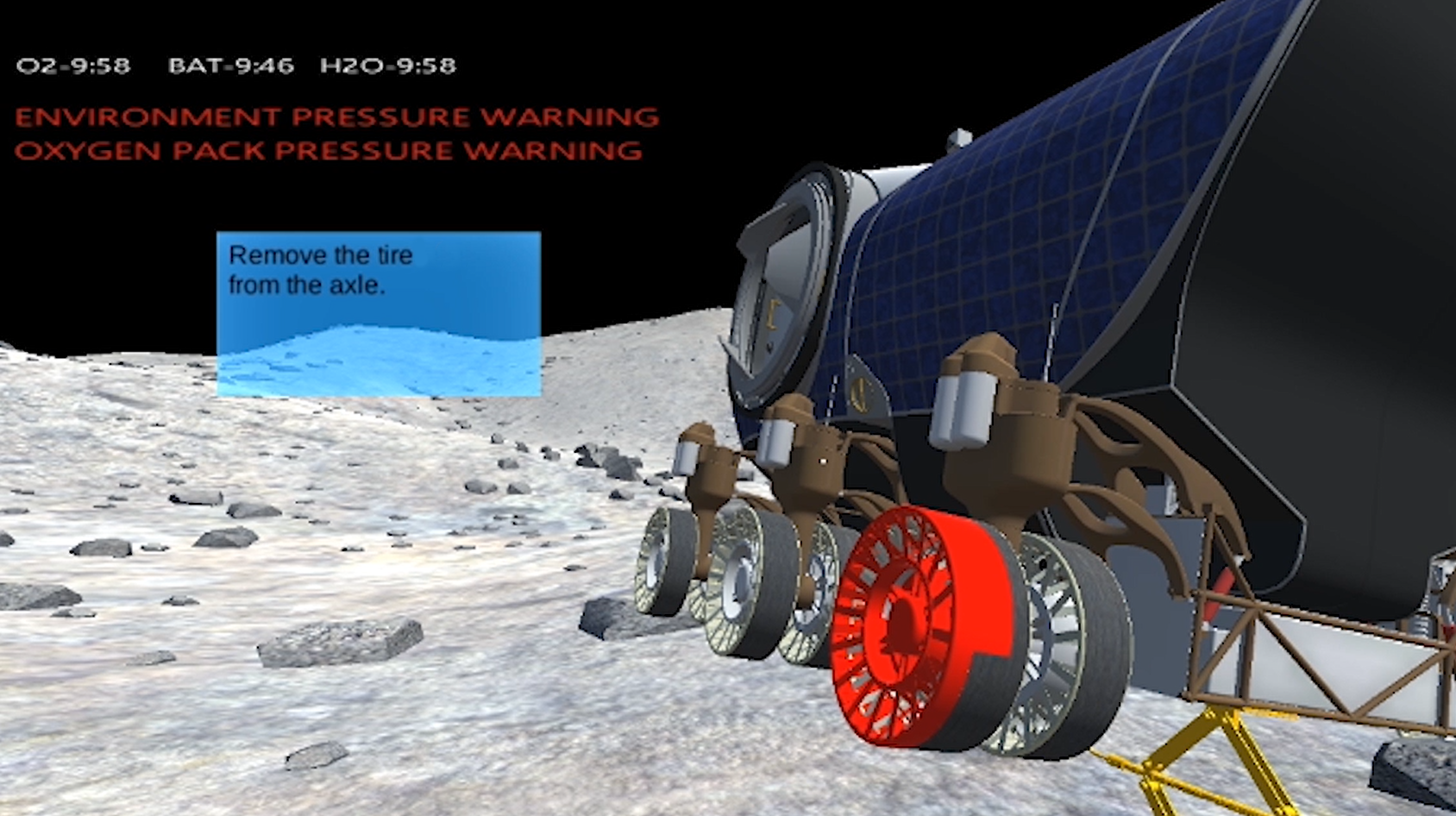}
\caption{Overlayed animation of removal of MMSEV Rover tire in red.}
\label{fig:tire}
\end{figure}

This allows for the system to understand the location and orientation of the rover so that instructional models may be inserted in reference to this model target. 

For the second requirement, the 6DOF pose is determined as a part of creating the task instructions.
These coordinates direct the animation of task sequences, through providing instructions on how to align the projected graphics with the model target.
Their sequence is predetermined, so the astronaut trainers would create packages of reference coordinates for task sequences prior to the mission, which could be rigorously tested as necessary on-site. 
For each step of the task, our system takes as input sets of Quaternion rotation values and position coordinates, which are given in relation to a model target located at the origin. 
It is important to mention the need to use Quaternion rotation values, as opposed to Euler angles, to avoid gimbal lock and ensure that the movement of models will execute correctly. With this information, the system can create moving overlays, or highlight specific areas of the rover at the origin, by providing one or a series of positions and rotations through which the projected model should appear. This ability is demonstrated in Figure \ref{fig:tire}, where the tire is overlaid onto the rover, demonstrating the process of its removal. 
Because we know the location of the model target, in this case the rover, and the location of a specific model relative to when this rover is at the origin, we can transform these coordinates. 
Through this transformation, the instructional models can be properly aligned with the model target; a bolt holding a tire in place on the target rover could be highlighted, or the process of removing the bolt could be projected onto the model. 

\subsection{Model Referential Communications}
Due to the limited bandwidth of deep-space transmission, a solution must be introduced to display and transfer models/instructions. This is particularly challenging given the average 3D model size is two or more orders of magnitude larger than the data size that can be transferred in a reasonable time frame (say 1--3 minutes) given the bandwidth speed available \cite{nasa_2019} and thus transferring or downloading models in real time is infeasible. 

As such, we take inspiration from multiplayer games and introduce a model-referential communication system wherein commonly used 3D models are stored locally on the deep-space device. To relay live AR instructions with interactive 3D models, the relative model position, rotation, and scale information is sent, as opposed to the whole model. This reduces the amount of information needed to be sent to just 11.8 KB for our entire instruction set, as opposed to over 54 MB for just a single rover model. This also increases the flexibility of instructions that can be sent from Mission Control: As long as no new 3D models are needed, new instructions or demonstrations can easily be sent within a matter of minutes. 

We implemented our model referential communication task using MongoDB \cite{mongodb} with Unity. We chose MongoDB over SQL \cite{microsoft} databases for our implementation, as it uses a document-based model suitable for instruction sets or 3D model collections. We created preexisting AssetBundles containing 3D models of commonly-used tools/vehicles stored offline on-device. Instruction sets were created in BSON \cite{bson} form with each instruction sub-step containing a dynamic link  to the required 3D models (stored on-device), as well as the required position and orientations for those models. When the instruction set (in BSON form) is received, the BSON-string is first serialized to a C\# object in Unity. We then find the corresponding AssetBundle(s) for the C\# object, along with the position and orientation for the given 3D models and project it onto the actual relative position and orientation of the rover in AR. This is used to provide step-by-step instructions projected onto the rover along with 3D demonstrations.

\begin{figure*}[!htb]
\centering
\includegraphics[width=0.9\textwidth]{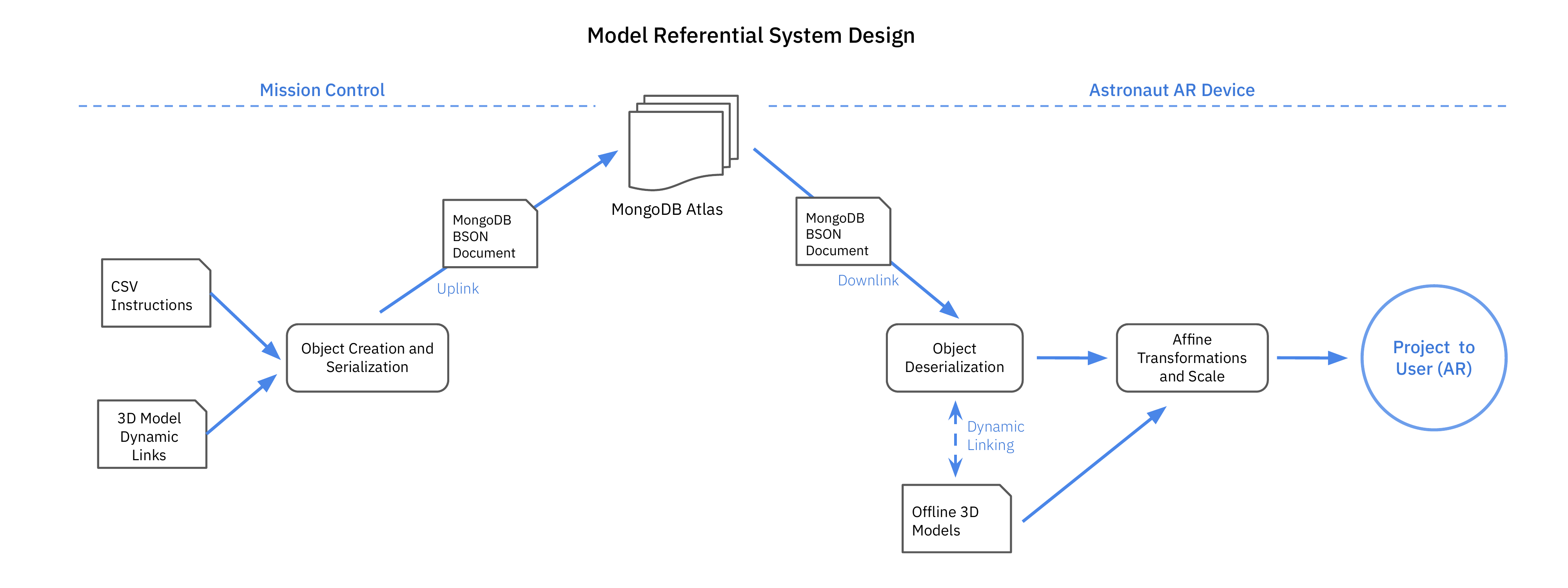}
\caption{Model Referential System Design}
\label{fig:ModelRefDesign}   
\end{figure*}

\begin{figure*}[!htb]
\centering
\includegraphics[width=0.9\textwidth]{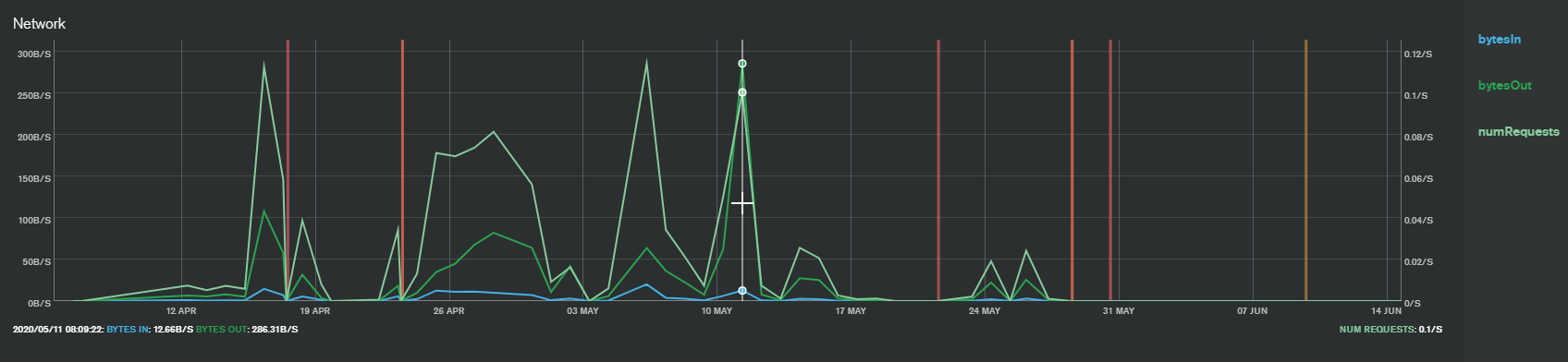}
\caption{Data/Bandwidth usage of our MongoDB Atlas Instance}
\label{fig:DBStats}   
\end{figure*}

For our proof of concept, we used a MongoDB Atlas cluster hosted on Amazon Web Services (AWS) \cite{AWScitation} to simulate a data downlink, as shown in Figure \ref{fig:ModelRefDesign}. We created a Java program that takes instructions in CSV format and automatically converts them into InstructionSet objects. The program also automatically deserializes these objects into MongoDB BSON document form and uploads them to our MongoDB Atlas instance. This simulates Mission Control sending new instruction sets to the astronauts via a data uplink. We were able to demonstrate the usability and feasibility of this approach both from a technical standpoint and from a limited bandwidth standpoint. As shown in Figure \ref{fig:DBStats}, the peak average bandwidth usage on a single day was only a mere 286 bytes/second, with other days clocking in at only 20--80 bytes/second. This is well within the bandwidth of the direct-to-Earth data rate in previous deep-space spacecraft (for example, the Curiosity rover had a direct-to-Earth data rate of 62.5-4000 bytes/sec). 

\section{Outcome and Lessons Learned}

\textit{Voice Control Implementation and Performance}: We were pleasantly surprised by the performance of our voice control implementation; updates were consistently available in five seconds or less. We found that short, simple commands seemed to work best, as they are easiest to recognize, and are intuitive for the user (e.g., \say{Open Instructions} and \say{Next}). Based upon our findings, a good direction for further work would be to pursue a method of interaction that is more natural, where the user can ask questions of the interface and expect responses, or move in nonlinear paths through tasks. 

\textit{AR Implementation and Testing}: Through the course of this project, we discovered it was helpful to develop AR aspects of this work in a manner compatible with VR (Virtual Reality) for testing purposes; we found that the unique challenges of the lunar surface were easier to mimic in VR. Additionally, we developed for a Microsoft HoloLens AR headset, but an interesting direction for future research would be to use eye tracking in combination with voice for navigation on an AR  headset capable of this, such as the Microsoft HoloLens 2.

\textit{Model Referential Task Instructions}: The model referential task instruction system we developed was very successful from both a technical and ease of use standpoint. A flight controller or Mission Control on Earth could easily write new instructions in CSV format using Microsoft Excel \cite{msexcel}, which can be automatically parsed, uploaded, and sent to the astronaut's AR device using the instruction parsing program we created. We found MongoDB BSON and MongoDB Atlas to be well suited to this task, as we could upload/download instructions in near real-time, and the MongoDB Unity driver worked well with our system. Finally, our model referential task instruction system offered huge savings from a data bandwidth perspective, as only the 6DOF poses of 3D models and their corresponding instructions were needed. Future work in this area would include a more seamless and user friendly way to upload instructions and also provide an intelligent mechanism for automatically scheduling data-heavy instructions during low-bandwidth usage periods. 

\textit{Object Recognition}: We were unable to evaluate the accuracy of our object recognition implementation, due to COVID-19 restricting travel to the location of the rover. Therefore, in the final stages of our work and testing, we used Vuforia Simulator Play Mode \cite{vuforiaplayMode}, Vuforia's tracking simulator, in order to test with our model targets. From our exploration, we would strongly recommend Vuforia as the best option for this type of task. We examined several other object recognition solutions, but Vuforia required the least work for the same performance. In particular, Vuforia offers an Object Scanner which can create model targets for training.

\textit{Sampling}:
Given the high accuracy of the voice recognition system for an experimental system with no specialized training (we used Microsoft Cognitive Services which has a Word Error Rate of $< 10\%$ \cite{microsoftaccuracy}), we recommend voice recording as a method for taking field notes in the future. In terms of visual data, we considered collecting videos and photos. However, file size is a concern with video capture, and video quality may not be ideal given the lighting conditions. The photo approach did not pose such problems, and we concluded that there would not be enough real-time changes in the environment to require constant visual feedback. Thus, we recommend the photo approach in order to minimize space usage and preserve visual data quality. Of course, if storage capacity of the final device permits, a hybrid solution of both photos and videos could be used for increased data fidelity. For future work, we believe there is potential to enhance the information conveyed by associating a sample with its location. Some suggestions are to use a predefined topographic map and then \say{pin} the location of the sample using some sort of visual marker.  Another option is to recreate the topography of a sample site using radar and LiDAR, and then display the field notes in the context of this generated site.

\section{Conclusions}
In this paper, we have presented an AR system with several design considerations for potential challenges faced in deep-space missions. Our contributions to this area are four-fold. First, we build upon prior art and introduce 3D graphical overlay instructions in AR that are designed to facilitate task completion and understanding by astronauts. We incorporate state-of-the-art object recognition technologies to allow our system to overlay graphics and animations with their associated instructions onto the target object directly. We hope that this combination of text and graphics increases astronauts' productivity and performance accuracy, especially given a foreign environment, novel instructions, and potentially poor lighting conditions. 

Second, we considered the physical limitations of space suits, and designed our AR  system with a voice control system. This aims to reduce astronaut fatigue, while providing a relatively reliable means of operation for non-critical tasks. By leveraging existing speech recognition models (with a Word Error Rate of less than 10\%), our system is able to reliably recognize instructions and commands. Furthermore, for tasks that require the astronaut to take notes of some form, we introduce a voice-based note-taking and compiling system, to reduce the need for a physical writing implement. 

Third, we introduce an unobtrusive, automatic telemetry monitoring system that will alert the astronauts to any anomalies and other serious issues. We hope that this system will allow astronauts to focus on the task-at-hand, without being worried about manually monitoring vitals or other telemetry streams. We believe this is especially important in deep-space missions, where the time delay between the astronaut and MCC will be too large to safely rely on MCC for anomalous telemetry monitoring. Finally, we present a model referential communications model to allow for an efficient method of transferring and linking large 3D data models (required for use in our graphical instructions) within the specification of deep-space data rate limitations. 

Our future work will focus on improvements on multiple fronts. First, we hope to expand our voice control system and introduce an intelligent assistant capable of handling more complex user requests without the need for cloud computing. Second, we intend to introduce an unsupervised machine learning based anomaly detection system, that will triage and cluster anomalies observed in telemetry data and provide helpful suggestions on solutions. Third, we aim to make our system more contextually aware from both a user standpoint and from an environmental standpoint (taking into account terrain, lighting, obstacles, and other data). 

\section{Acknowledgments}

This work would not have been possible without the effort and advice of Professor Steve Feiner. We are extremely grateful for his support and enthusiasm. 
Additionally, we would like to extend our sincerest gratitude to the Columbia University Fu Foundation School of Engineering and Applied Science, for academic and financial support, the NASA SUITS Challenge Program for providing us with a platform to create and share our work, and The Columbia Space Initiative, for their constant encouragement and tireless passion.

\bibliographystyle{IEEEtran}
\interlinepenalty=10000
\bibliography{refs}   

\begin{thebibliography}{10}
\providecommand{\url}[1]{#1}
\csname url@samestyle\endcsname
\providecommand{\newblock}{\relax}
\providecommand{\bibinfo}[2]{#2}
\providecommand{\BIBentrySTDinterwordspacing}{\spaceskip=0pt\relax}
\providecommand{\BIBentryALTinterwordstretchfactor}{4}
\providecommand{\BIBentryALTinterwordspacing}{\spaceskip=\fontdimen2\font plus
\BIBentryALTinterwordstretchfactor\fontdimen3\font minus
  \fontdimen4\font\relax}
\providecommand{\BIBforeignlanguage}[2]{{%
\expandafter\ifx\csname l@#1\endcsname\relax
\typeout{** WARNING: IEEEtran.bst: No hyphenation pattern has been}%
\typeout{** loaded for the language `#1'. Using the pattern for}%
\typeout{** the default language instead.}%
\else
\language=\csname l@#1\endcsname
\fi
#2}}
\providecommand{\BIBdecl}{\relax}
\BIBdecl

\bibitem{LOVE2013}
\BIBentryALTinterwordspacing
S.~G. Love and M.~L. Reagan, ``Delayed voice communication,'' \emph{Acta
  Astronautica}, vol.~91, pp. 89 -- 95, 2013. [Online]. Available:
  \url{http://www.sciencedirect.com/science/article/pii/S0094576513001537}
\BIBentrySTDinterwordspacing

\bibitem{PATTERSON1999}
E.~Patterson, J.~Watts‐Englert, and D.~Woods, ``Voice loops as coordination
  aids in space shuttle mission control,'' \emph{Computer supported cooperative
  work : CSCW : an international journal}, vol.~8, pp. 353--71, 02 1999.

\bibitem{DUNBAR2020}
B.~Dunbar, \emph{Moon to Mars Overview}, 2020,
  \url{https://www.nasa.gov/topics/moon-to-mars/overview}.

\bibitem{mitra2018human}
P.~Mitra, \emph{Human Systems Integration of an Extravehicular Activity Space
  Suit Augmented Reality Display System}.\hskip 1em plus 0.5em minus
  0.4em\relax Mississippi State University, 2018.

\bibitem{feinerpaper2}
S.~J. {Henderson} and S.~K. {Feiner}, ``Augmented reality in the psychomotor
  phase of a procedural task,'' in \emph{Proc. 10th IEEE International
  Symposium on Mixed and Augmented Reality}, 2011, pp. 191--200.

\bibitem{OLSSON2012}
T.~Olsson, T.~Kärkkäinen, E.~Lagerstam, and L.~Ventä-Olkkonen, ``User
  evaluation of mobile augmented reality scenarios,'' \emph{JAISE}, vol.~4, pp.
  29--47, 01 2012.

\bibitem{stevepaper}
M.~Baltrusaitis and K.~Feigh, ``The development of a user interface for
  mixed-initiative plan management for human spaceflight,'' in \emph{2019 IEEE
  Aerospace Conference}.\hskip 1em plus 0.5em minus 0.4em\relax IEEE, 2019, pp.
  1--9.

\bibitem{dsouza}
G.~V. D'souza, ``The effectiveness of augmented reality for astronauts on lunar
  missions: An analog study,'' Master's thesis, Embry-Riddle Aeronautical
  University, 1 Aerospace Boulevard Daytona Beach, FL 32114-3900, 12 2019.

\bibitem{arpaper}
\BIBentryALTinterwordspacing
R.~T. Azuma, ``A survey of augmented reality,'' \emph{Presence: Teleoperators
  and Virtual Environments}, vol.~6, no.~4, pp. 355--385, 1997. [Online].
  Available: \url{https://doi.org/10.1162/pres.1997.6.4.355}
\BIBentrySTDinterwordspacing

\bibitem{pokemongo}
\BIBentryALTinterwordspacing
``Pokémon go!'' [Online]. Available: \url{https://www.pokemongo.com/en-us/}
\BIBentrySTDinterwordspacing

\bibitem{ingress}
\BIBentryALTinterwordspacing
``Ingress.'' [Online]. Available: \url{https://www.ingress.com/}
\BIBentrySTDinterwordspacing

\bibitem{MARS2018}
K.~Mars, \emph{Deep Space}, 2018,
  \url{https://www.nasa.gov/johnson/exploration/deep-space}.

\bibitem{BRALY2019}
\BIBentryALTinterwordspacing
A.~M. Braly, B.~Nuernberger, and S.~Y. Kim, ``Augmented reality improves
  procedural work on an international space station science instrument,''
  \emph{Human Factors}, vol.~61, no.~6, pp. 866--878, 2019, pMID: 30694084.
  [Online]. Available: \url{https://doi.org/10.1177/0018720818824464}
\BIBentrySTDinterwordspacing

\bibitem{HELIN2018}
\BIBentryALTinterwordspacing
K.~Helin, T.~Kuula, C.~Vizzi, J.~Karjalainen, and A.~Vovk, ``User experience of
  augmented reality system for astronaut's manual work support,''
  \emph{Frontiers in Robotics and AI}, vol.~5, 2018. [Online]. Available:
  \url{https://doi.org/10.3389/frobt.2018.00106}
\BIBentrySTDinterwordspacing

\bibitem{MARKOV2013}
D.~Markov-Vetter and O.~Staadt, ``A pilot study for augmented reality supported
  procedure guidance to operate payload racks on-board the international space
  station,'' in \emph{2013 IEEE International Symposium on Mixed and Augmented
  Reality (ISMAR)}.\hskip 1em plus 0.5em minus 0.4em\relax IEEE, 2013, pp.
  1--6.

\bibitem{ramsey_2015}
\BIBentryALTinterwordspacing
S.~Ramsey, ``Nasa microsoft collaborate,'' Jun 2015. [Online]. Available:
  \url{https://www.nasa.gov/press-release/nasa-microsoft-collaborate-to-bring-science-fiction-to-science-fact}
\BIBentrySTDinterwordspacing

\bibitem{KINTZCHOU2016}
\BIBentryALTinterwordspacing
N.~M. Kintz, C.-P. Chou, W.~B. Vessey, L.~B. Leveton, and L.~A. Palinkas,
  ``Impact of communication delays to and from the international space station
  on self-reported individual and team behavior and performance: A
  mixed-methods study,'' \emph{Acta Astronautica}, vol. 129, pp. 193 -- 200,
  2016. [Online]. Available:
  \url{http://www.sciencedirect.com/science/article/pii/S009457651630697X}
\BIBentrySTDinterwordspacing

\bibitem{TAYLOR2014}
N.~J. Jim~Taylor, \emph{Deep Space Communications}, 2014,
  \url{https://descanso.jpl.nasa.gov/monograph/series13/DeepCommoOverall--141030A_ama.pdf}.

\bibitem{AzureCogServices}
\BIBentryALTinterwordspacing
``Cognitive services.'' [Online]. Available:
  \url{https://azure.microsoft.com/en-us/services/cognitive-services/}
\BIBentrySTDinterwordspacing

\bibitem{WONG2017}
C.~Wong, \emph{The Saga of Writing in Space}, 2017,
  \url{https://airandspace.si.edu/stories/editorial/saga-writing-space}.

\bibitem{technologies}
\BIBentryALTinterwordspacing
U.~Technologies. [Online]. Available: \url{https://unity.com/}
\BIBentrySTDinterwordspacing

\bibitem{MCGREGOR2013}
C.~{McGregor}, ``A platform for real-time online health analytics during
  spaceflight,'' in \emph{2013 IEEE Aerospace Conference}, 2013, pp. 1--8.

\bibitem{feiner1993knowledge}
S.~Feiner, B.~Macintyre, and D.~Seligmann, ``Knowledge-based augmented
  reality,'' \emph{Communications of the ACM}, vol.~36, no.~7, pp. 53--62,
  1993.

\bibitem{vuforia}
\BIBentryALTinterwordspacing
``Vuforia.'' [Online]. Available: \url{https://developer.vuforia.com/}
\BIBentrySTDinterwordspacing

\bibitem{nasa_2019}
\BIBentryALTinterwordspacing
``Communications with earth,'' Aug 2019. [Online]. Available:
  \url{https://mars.nasa.gov/msl/mission/communications/}
\BIBentrySTDinterwordspacing

\bibitem{mongodb}
\BIBentryALTinterwordspacing
``Mongodb.'' [Online]. Available: \url{https://www.mongodb.com/}
\BIBentrySTDinterwordspacing

\bibitem{microsoft}
\BIBentryALTinterwordspacing
``Microsoft sql.'' [Online]. Available:
  \url{https://www.microsoft.com/en-us/sql-server}
\BIBentrySTDinterwordspacing

\bibitem{bson}
\BIBentryALTinterwordspacing
``Bson.'' [Online]. Available: \url{http://bsonspec.org/}
\BIBentrySTDinterwordspacing

\bibitem{AWScitation}
\BIBentryALTinterwordspacing
M.~Hunt, ``Aws free tier,'' 1994. [Online]. Available:
  \url{https://aws.amazon.com/free/?all-free-tier.sort-by=item.additionalFields.SortRank&all-free-tier.sort-order=asc}
\BIBentrySTDinterwordspacing

\bibitem{msexcel}
\BIBentryALTinterwordspacing
{Microsoft Corporation}, ``Microsoft excel.'' [Online]. Available:
  \url{https://office.microsoft.com/excel}
\BIBentrySTDinterwordspacing

\bibitem{vuforiaplayMode}
\BIBentryALTinterwordspacing
``Vuforia play mode.'' [Online]. Available:
  \url{https://library.vuforia.com/content/vuforia-library/en/articles/Solution/vuforia-play-mode-in-unity.html}
\BIBentrySTDinterwordspacing

\bibitem{microsoftaccuracy}
\BIBentryALTinterwordspacing
``Evaluate custom speech accuracy.'' [Online]. Available:
  \url{https://docs.microsoft.com/en-us/azure/cognitive-services/speech-service/how-to-custom-speech-evaluate-data}
\BIBentrySTDinterwordspacing

\end{thebibliography}
\end{document}